\documentclass[a4paper]{article}

\usepackage{INTERSPEECH2020}
\usepackage{cite}
\usepackage{multirow}
\usepackage{adjustbox}

\title{Improving Multi-Scale Aggregation Using Feature Pyramid Module\\for Robust Speaker Verification of Variable-Duration Utterances}
\name{Youngmoon Jung, Seong Min Kye, Yeunju Choi, Myunghun Jung, Hoirin Kim}
\address{School of Electrical Engineering, KAIST, Daejeon, South Korea}
\email{\{dudans,kye9165,wkadldppdy,kss2517,hoirkim\}@kaist.ac.kr}

\begin{document}

\maketitle
\begin{abstract}
Currently, the most widely used approach for speaker verification is the deep speaker embedding learning. In this approach, 
%convolutional neural networks are mainly used as a speaker feature extractor, and 
we obtain a speaker embedding vector by pooling single-scale features that are extracted from the last layer of a speaker feature extractor.
%we extract speaker embeddings from the last layer of the frame-level feature extractor such as a convolutional neural network (CNN).
Multi-scale aggregation (MSA), which utilizes multi-scale features from different layers of the feature extractor, has recently been introduced and shows superior performance for variable-duration utterances.
%shows better duration-robustness.
%has shown superior performance for both short and long utterances.
% from different time-frequency resolutions
%hierarchical time-frequency context information, 
%has recently achieved significant performance improvement for the deep speaker embedding framework. 
%over the conventional deep speaker embedding which only uses a last feature map. 
To increase the robustness dealing with utterances of arbitrary duration, this paper improves the MSA by using a feature pyramid module. The module enhances speaker-discriminative information of features from multiple layers via a top-down pathway and lateral connections. 
%To extract speaker embeddings, we combine the resulting features containing rich speaker information at different resolutions. 
%Then, we extract speaker embeddings using the resulting features which contain rich speaker information at different resolutions.
We extract speaker embeddings using the enhanced features that contain rich speaker information with different time scales. %time-frequency scales.
Experiments on the VoxCeleb dataset show that the proposed module improves previous MSA methods with a smaller number of parameters. It also achieves better performance than state-of-the-art approaches for both short and long utterances. 
\end{abstract}
\noindent\textbf{Index Terms}: Speaker verification, deep speaker embedding, multi-scale aggregation, feature pyramid module, short duration

\section{Introduction}
\label{sec:intro}

Speaker verification (SV) is the task of verifying a speaker's claimed identity based on his or her speech. 
Depending on the constraint of the transcripts used for enrollment and verification, SV systems fall into two categories, text-dependent SV (TD-SV) and text-independent SV (TI-SV). TD-SV requires the content of input speech to be fixed, while TI-SV operates on unconstrained speech.
Before the deep learning era, the combination of \textit{i}-vector \cite{Dehak2011} and probabilistic linear discriminant analysis (PLDA) \cite{Ioffe2006} was the dominant approach for SV \cite{Kenny2010, Garcia2011}. 
Although this approach performs well with long utterances, it suffers from performance degradation with short utterances.

Recently, the most widely used SV approach is 
%to extract speaker embeddings directly from a speaker discriminative network 
the deep speaker embedding learning which extracts speaker embeddings directly from a deep-learning-based speaker-discriminative network \cite{Li2017, Zhang2018, Nagrani2017, Chung2018, Cai2018, Snyder2018, Okabe2018, Tang2019, Gao2019, Seo2019, Hajavi2019, Jung2019, Jung2019asru, Kye2020}. %In this framework, convolutional neural networks (CNNs) such as ResNet \cite{Li2017, Chung2018, Cai2018, Jung2019, Gao2019, Seo2019, Jung2019asru, Hajavi2019} or time-delay neural network (TDNN) \cite{Snyder2018, Okabe2018, Tang2019} are mostly used as the speaker-discriminative network.
This approach outperforms \textit{i}-vector/PLDA approach, especially on short utterances.
In deep speaker embedding learning, convolutional neural networks (CNNs) such as time-delay neural network (TDNN) \cite{Snyder2018, Okabe2018, Tang2019} or ResNet \cite{Li2017, Chung2018, Cai2018, Jung2019, Gao2019, Seo2019, Jung2019asru, Hajavi2019, Kye2020, JungMH2020} are mostly used as the speaker-discriminative network.
% , a special type of 1D-CNN
%deep residual network (ResNet)
%In such systems, DNNs are employed to learn speaker embeddings, termed d-vectors. Similarly to as in the case of i-vectors, such embedding can then be used to represents utterances in a fixed dimensional space. 
Specifically, the network is trained to classify training speakers \cite{Cai2018, Snyder2018, Okabe2018, Jung2019, Gao2019, Seo2019, Jung2019asru, Tang2019, Hajavi2019, Kye2020, JungMH2020} or to separate same-speaker and different-speaker utterance pairs  \cite{Li2017, Zhang2018}.
After training, an utterance-level speaker embedding called deep speaker embedding is obtained by aggregating speaker features extracted from the network.
% using cross-entropy loss
% using triplet loss
%convolutional neural networks (CNNs) such as ResNet \cite{Li2017, Chung2018, Cai2018, Jung2019, Gao2019, Seo2019, Jung2019asru, Hajavi2019} and time-delay neural network (TDNN) \cite{Snyder2018, Okabe2018, Tang2019} which is a special case of 1-D CNN.
Most of these approaches use a pooling layer for feature aggregation, mapping variable-length speaker features to a fixed-dimensional embedding. There are several pooling methods, such as global average pooling \cite{Li2017, Nagrani2017, Seo2019}, statistics pooling \cite{Snyder2018, Gao2019}, attentive statistics pooling \cite{Okabe2018}, learnable dictionary encoding \cite{Cai2018}, and spatial pyramid encoding \cite{Jung2019}. 

Meanwhile, all these pooling layers use only single-scale features from the last layer of the feature extractor. 
To aggregate speaker information from different time scales, multi-scale aggregation (MSA) methods have been proposed recently \cite{Gao2019, Seo2019, Tang2019, Hajavi2019}. The MSA aggregates multi-scale features from different layers of a feature extractor to generate a speaker embedding. In \cite{Gao2019, Seo2019, Tang2019, Hajavi2019}, the authors show the effectiveness of the MSA in dealing with short or long utterances.
%More details are provided in Section \ref{sec:priorworks}.
% 그들은 last single feature만 이용하는 것보다 다양한 level에서 정보를 취합하여 embedding 추출하는 것이 더 좋은 성능을 보인다는 것을 밝혔다.

In this work, we propose a new MSA method using a feature pyramid module (FPM).
%for deep speaker embedding systems. 
% ResNet 기반의 feature extractor에 대해 multi-level aggregation 방식을 도입하며, 여기에 top-down pathway와 lateral connection을 포함한 feature pyramid module을 추가한다. 이로 인해 비슷한 파라미터 수를 유지하며, voxceleb DB에 대해 더 좋은 성능을 얻는다. 제안한 multi-level aggregation에 TAP, LDE, SAP의 3가지 pooling layer를 적용하여 각각의 pooling layer에 대해서 성능 향상이 있음을 보인다. 
A top-down architecture with lateral connections is used to generate feature maps with rich speaker information at all selected layers. Then, we exploit the rich multi-scale features of a ResNet-based feature extractor to extract speaker embeddings. 
In addition, we present a novel interpretation of the MSA using the theory of \cite{Veit2016}.
We evaluate our method using various pooling layers for TI-SV on the VoxCeleb dataset \cite{Nagrani2017, Chung2018}. Experimental results show that the performance of MSA is further improved by the FPM with a smaller number of parameters. Besides, the effectiveness of our method is verified on variable-duration test utterances.
%with the top-down architecture.
% 실험 관련 얘기 : 제안한 multi-level aggregation에 TAP, LDE, SAP의 3가지 pooling layer를 적용하여 각각의 pooling layer에 대해서 성능 향상이 있음을 보인다. 

%The rest of the paper is organized as follows. In Section 2, we discuss the relation between our work and the previous approaches. Section 3 presents our proposed method. The experimental setup and results are described in Section 4 and Section 5, respectively. We conclude our work in Section 6.

\section{Relation to prior works}
\label{sec:priorworks}

%In this section, we review prior works on multi-level aggregation and discuss the relation between our work and previous approaches.
Gao \textit{et al.} \cite{Gao2019} proposed multi-stage aggregation, where ResNet is used as a feature extractor. 
The output feature maps of stage 2, 3, and 4 (see Table \ref{architecture}) are concatenated along the channel axis.
%into a single feature map. 
%Three feature maps of the last layers in stage 2, 3, and 4 (refer to Table \ref{architecture}) are concatenated into a single feature map. 
To make feature maps match in time-frequency resolution, 
the feature map from stage 2 is downsampled by convolution with stride 2, and the feature map from stage 4 is upsampled by bilinear interpolation or transposed convolution. 
%Then, statistics pooling is applied to the concatenated features.
After concatenation, speaker embeddings are generated by statistics pooling.
%After concatenation, statistics pooling is applied to generate speaker embeddings. 
%By exploiting hierarchical time-frequency context information, they achieved state-of-the-art performance on VoxCeleb1.

Seo \textit{et al.} \cite{Seo2019} also utilize feature maps from different stages of ResNet to fuse information at different resolutions.
Unlike the method of Gao \textit{et al.}, global average pooling (GAP) is applied to the feature maps, and the pooled feature vectors are concatenated into a long vector.
The concatenated vector is fed into fully-connected layers to generate the speaker embedding.
Hajavi \textit{et al.} \cite{Hajavi2019} proposed a similar approach to the study of Seo \textit{et al.} Their proposed model, UtterIdNet, shows significant improvement in speaker recognition with short utterances.

\begin{figure}[t]
  %\vspace{-0.05cm}
  \centerline{\includegraphics[width=6.8cm]{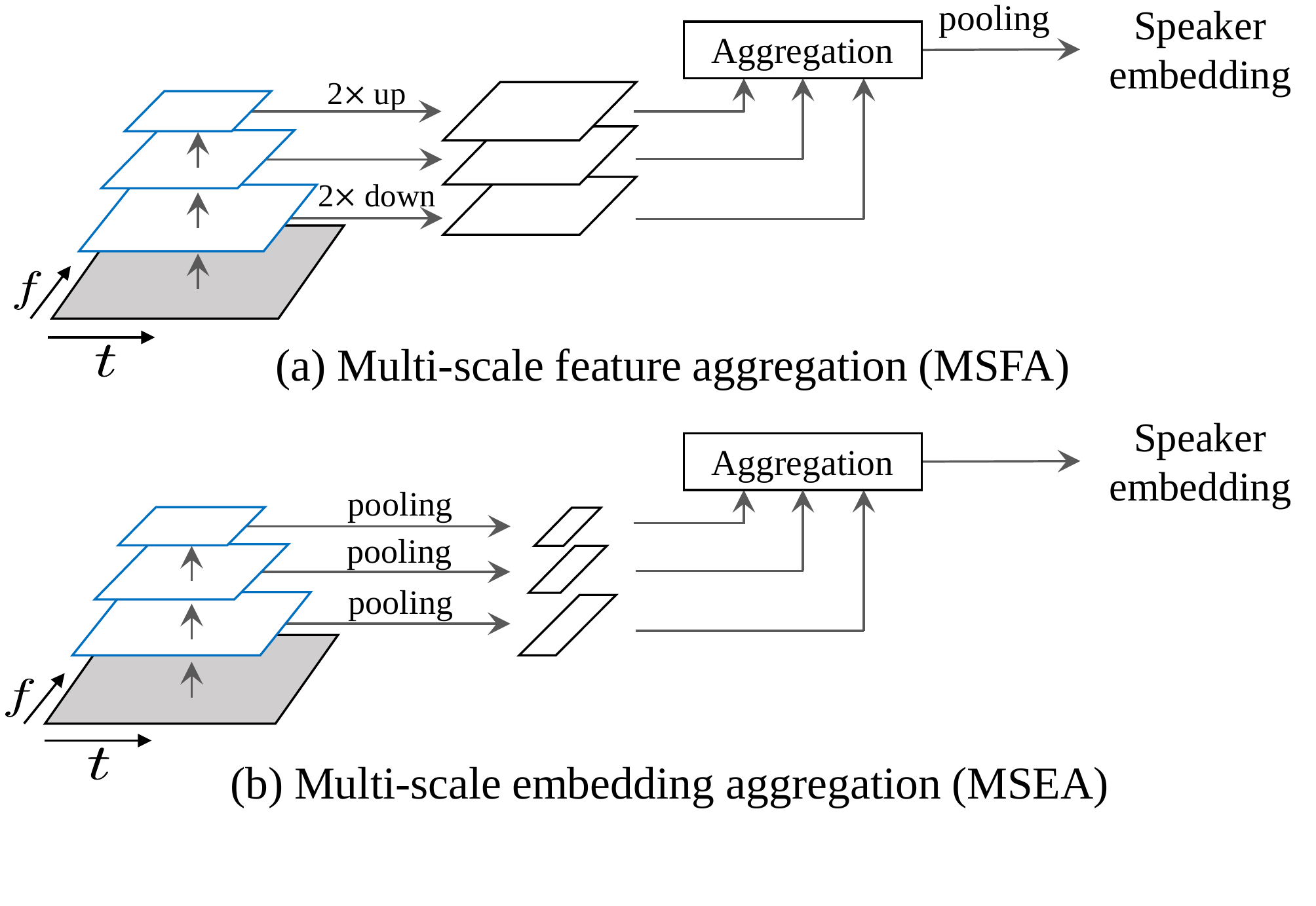}}
  \vspace{-0.65cm}
  \caption{Two types of multi-scale aggregation (MSA). (a) Multi-scale feature aggregation (MSFA). (b) Multi-scale embedding aggregation (MSEA). In this paper, acoustic features of consecutive frames are indicated by grey rectangles and feature maps are indicated by blue outlines. ``2$\times$up (or down)" is an upsampling (or downsampling) by a factor of 2.}
  \vspace{-0.21cm}
  \label{MSFA_MSEA}
\end{figure}

To sum up, the method of Gao \textit{et al.} fuses feature maps from different stages to form a single feature map, and applies pooling to the fused feature map to generate the speaker embedding. On the other hand, the method of Seo \textit{et al.} applies pooling to feature maps respectively, and aggregates the resulting vectors to obtain the speaker embedding. In this paper, we take these two methods as baselines. For clarity, we denote the first method as multi-scale feature aggregation (MSFA) and the second one as multi-scale embedding aggregation (MSEA). These approaches are illustrated in Fig. \ref{MSFA_MSEA}. We will show that the proposed feature pyramid module improves both baselines. 

Meanwhile, these studies show the effectiveness of MSA on only one type of pooling operation, i.e., statistics pooling and GAP, respectively. Unlike these studies, we evaluate our approach using three popular pooling methods: GAP, self-attentive pooling (SAP) \cite{Cai2018}, and learnable dictionary encoding (LDE) \cite{Cai2018}.
Experimental results show that the proposed method achieves good performance for the three pooling layers.

%In another related work, Hajavi \textit{et al.} \cite{Hajavi2019} proposed a similar approach to the study of Seo \textit{et al.}
%They extract three embeddings from different identity blocks.
%Then, a non-linear aggregator combines the embeddings to produce a speaker representation.
%Their proposed model, UtterIdNet, shows significant improvement in speaker recognition with short speech segments.

%Tang \textit{et al.} \cite{Tang2019} proposed a hybrid neural network structure where long short-term memory (LSTM) layers are added on top of TDNN layers. They proposed a multi-level pooling strategy to fuse different speaker information from both TDNN and LSTM layers.

\section{Proposed approach}

In this section, we introduce the proposed 
multi-scale aggregation using the feature pyramid module motivated by \cite{Lin2017}. 
%As we explained in the introduction, CNN-based feature extractor is widely used in modern speaker verification systems. 
In this work, we use ResNet as our feature extractor just as in the baselines \cite{Gao2019, Seo2019}. The architecture is described in Table \ref{architecture}. 
%The ResNet takes 64-dimensional log Mel-filterbank (Fbank) features of $T$ frames in each utterance.

\subsection{Multi-scale aggregation}

Deep CNNs such as ResNet are usually bottom-up, feed-forward architectures, which use repeated convolutional and subsampling layers to learn sophisticated feature representations.
%for the input.
Deep CNNs compute a feature hierarchy layer by layer, and with subsampling layers, the feature hierarchy is inherently multi-scale of pyramidal shape.
This in-network feature hierarchy produces feature maps of different time-frequency scales and resolutions, but introduces large semantic gaps caused by different depths. 
%Feature maps of higher layers capture more semantic information but have lower resolutions due to the repeated pooling in the lower layers.
%From the speaker recognition point of view, acoustic features of consecutive frames is the input. 
%In deep speaker embedding systems, 
In SV, as the network is trained to discriminate speakers, features of higher layers contain more speaker-discriminative information (higher-level speaker information) 
but have lower resolutions due to the repeated subsampling.
%in lower layers.

\begin{figure}[t]
  %\vspace{0.3cm}
  \centerline{\includegraphics[width=7.3cm]{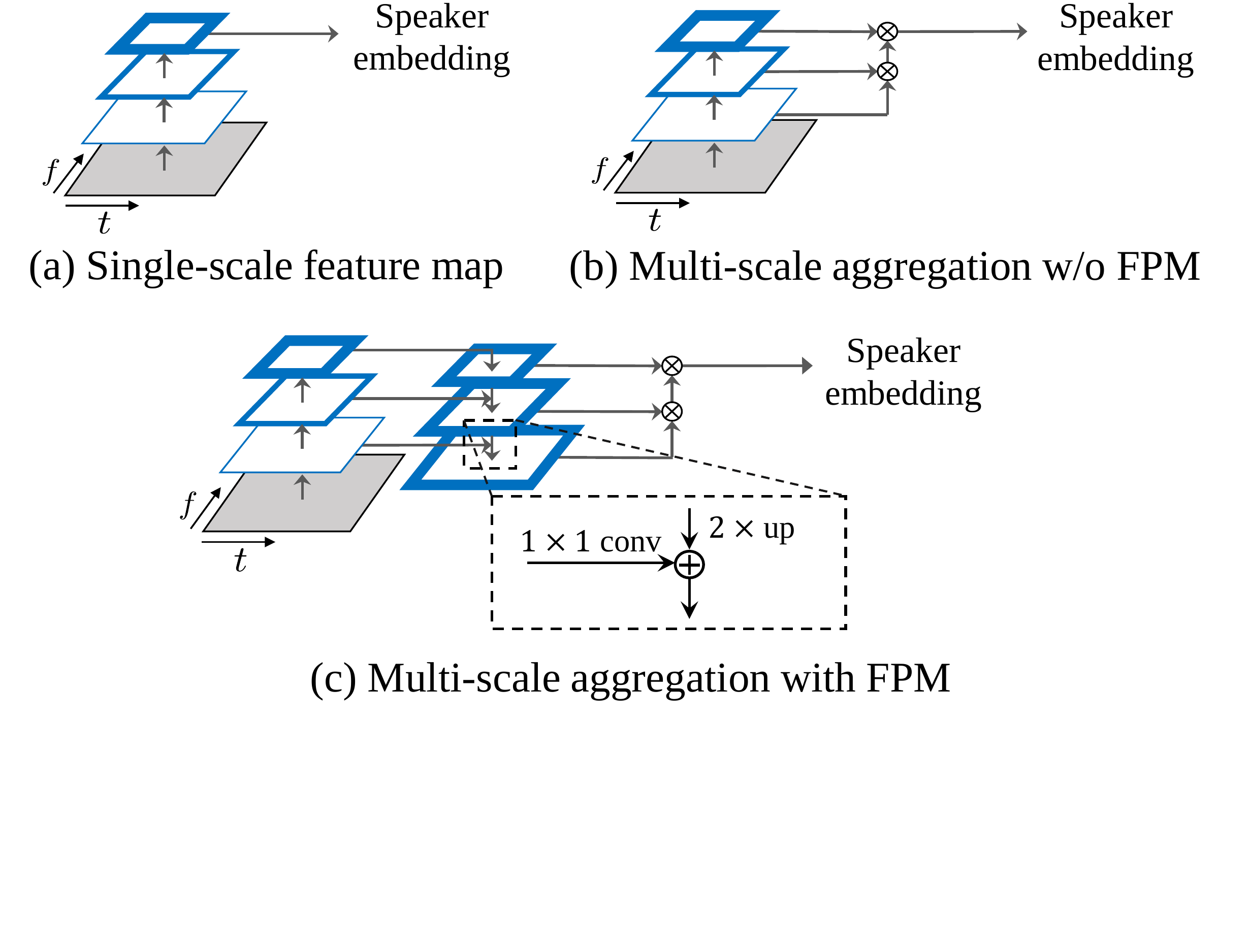}}
  \vspace{-1.65cm}
  \caption{How to use feature maps for deep speaker embedding. (a) Using only single-scale feature maps. (b) Using multi-scale feature maps without feature pyramid module (FPM). (c) Using multi-scale feature maps with FPM. In this paper, thicker outlines denote feature maps with more speaker-discriminative information. $\otimes$ : concatenation, $\oplus$ : element-wise addition.}
  \vspace{-0.5cm}
  \label{PFH_FPN}
\end{figure}

\begin{table}[t]
\centering
%\begin{small}
\begin{scriptsize}
%\end{footnotesize}
\renewcommand{\arraystretch}{1.1}
\caption{The architecture of the feature extractor based on ResNet-34 \cite{He2016}. 
Inside the brackets is the shape of a residual block and outside the brackets is the number of stacked blocks on a stage.
The input size is $64 \times T$.}
%, where 64 is the feature dimension and T is the number of frames in each segment.}
\vspace{-0.27cm}
\label{architecture}
%\begin{footnotesize}
\begin{tabular}{c|c|c|cll}
%\begin{tabular}{|*3{>{\renewcommand{\arraystretch}{1}}c|}}
\cline{1-4} 
Layer name & Output size    & ResNet-34   & Stage  \\ \cline{1-4}
conv1      & $64 \times T \times 32$    & $7 \times 7, 32$, stride $1$    & -                                          \\ \cline{1-4}
conv2\_x       & $64 \times T \times 32$    & $\left[ \begin{array}{cc} 3 \times 3, 32  \\ 3 \times 3, 32 \end{array}\right]$ $\times$ $3$ & 1 \\ \cline{1-4}
conv3\_x       & $32 \times T/2 \times 64$  & $\left[ \begin{array}{cc} 3 \times 3, 64  \\ 3 \times 3, 64 \end{array}\right]$ $\times$ $4$ & 2 \\ \cline{1-4}
conv4\_x       & $16 \times T/4 \times 128$ & $\left[ \begin{array}{cc} 3 \times 3, 128  \\ 3 \times 3, 128 \end{array}\right]$ $\times$ $6$ & 3 \\ \cline{1-4}
conv5\_x       & $8 \times T/8 \times 256$  & $\left[ \begin{array}{cc} 3 \times 3, 256  \\ 3 \times 3, 256 \end{array}\right]$ $\times$ $3$ & 4 \\ \cline{1-4}
\cline{1-4}

\end{tabular}
%\end{small}
\end{scriptsize}
\vspace{-0.35cm}
\end{table}

%Aside from being capable of representing %higher-level semantics, 
%higher-level speaker information, 
Deep CNNs are robust to variance in scale and thus facilitate extraction of speaker embeddings from feature maps computed on a single input scale (Fig. \ref{PFH_FPN}(a)). However, even with this robustness, using multi-scale features from multiple layers (Fig. \ref{PFH_FPN}(b)) improves the performance as discussed in Section \ref{sec:priorworks}.

According to the previous works, the MSA improves speaker recognition performance by extracting speaker embeddings from multiple temporal scales \cite{Gao2019, Tang2019}.
%According to the previous works, the advantage of MSA is that it extracts speaker embeddings from multiple temporal scales, improving speaker recognition performance \cite{Gao2019, Tang2019}. 
%. They have shown that using multiple temporal scales improves speaker recognition performance \cite{Gao2019, Tang2019, Wang2019}. 
It is also useful for short-duration speaker recognition through an efficiently increased use of information in short utterances \cite{Hajavi2019}. 
Besides, it passes error signals back to earlier layers, which helps alleviate the vanishing gradient problem \cite{Tang2019, Seo2019}.

Two types of MSA are described in Fig. \ref{MSFA_MSEA}. For MSFA (Fig. \ref{MSFA_MSEA}(a)), we use the same method in \cite{Gao2019}. For MSEA (Fig. \ref{MSFA_MSEA}(b)), $1\times1$ convolutions are added before pooling.
The embeddings of different stages are concatenated and the output of the following fully-connected (FC) layer is used as the speaker embedding.

%Another view of MSA is 

%A deep CNN computes a feature hierarchy layer by layer, and with subsampling layers the feature hierarchy has an inherent multiscale, pyramidal shape. This in-network feature hierarchy produces feature maps of different resolutions, but introduces large differences in speaker information caused by different depths.
\subsection{Feature pyramid module}
\label{sec:FPM_section}

In deep CNNs, feature maps of lower layers have less speaker-discriminative information compared to those of higher layers. 
%The low-level speaker features harm their representational capacity for deep speaker embedding.
Intuitively, if we can enhance the speaker discriminability of the low-layer feature maps, the performance of MSA will be improved. Motivated by this intuition, we aim to create multi-scale features that have high-level speaker information at all layers.
%The goal of this paper is to create multi-level features that have rich speaker information at all levels.
The feature pyramid module (FPM) is used to achieve this goal. 
%This module consists of a top-down pathway and lateral connections. 
The MSA with FPM is illustrated in Fig. \ref{PFH_FPN}(c). 
The dotted box indicates the building block of FPM that consists of the lateral connection and the top-down pathway, merged by addition.

\begin{figure}[t]
  %\vspace{-0.1cm}
  \centerline{\includegraphics[width=7.9cm]{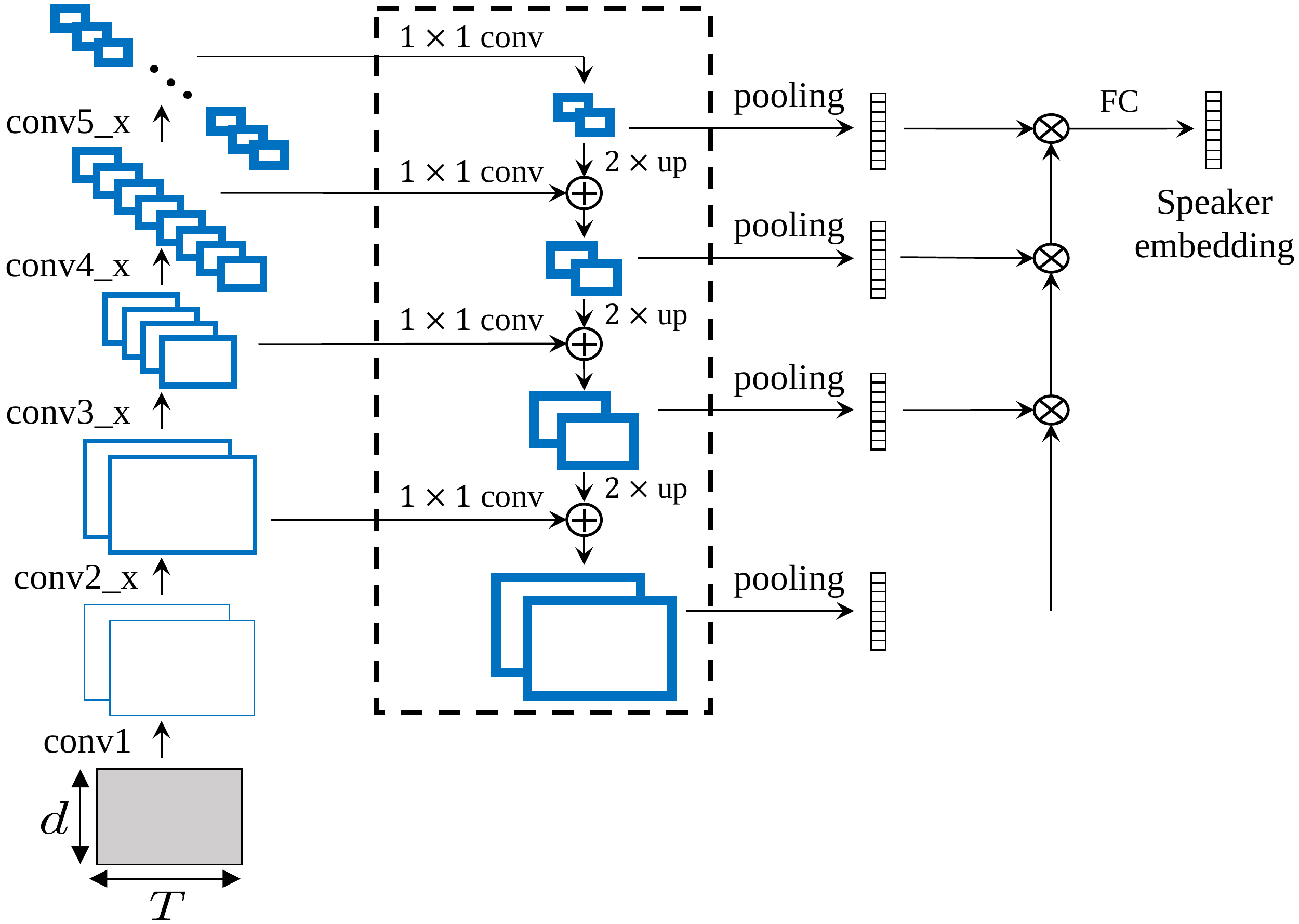}}
  \vspace{-0.2cm}
  \caption{Proposed MSA with FPM. The black dotted box indicates the FPM. Only MSEA is illustrated here, but the FPM can be applied to both MSFA and MSEA.}
  \vspace{-0.42cm}
  \label{PFH_FPN_details}
\end{figure}

The detailed architecture is shown in Fig. \ref{PFH_FPN_details}, involving a bottom-up pathway, a top-down pathway, and lateral connections. The bottom-up pathway is the feed-forward computation of the backbone ResNet. It computes a feature hierarchy consisting of feature maps at multiple scales with a scaling step of 2. 
%To achieve this goal, we use the feature pyramid module (FPM) that consists of a top-down pathway and lateral connections (Fig. \ref{PFH_FPN}(c)).
In each ResNet stage, there are many layers producing feature maps of the same scale (see Table \ref{architecture}). 
We select the last layer among the multiple layers of a stage, and then use the output of the layer as the output of the stage.
It is because the deepest layer contains the strongest features.
%as the output of the stage, since the deepest layer contains the strongest features.
%We choose the output of the last layer of a stage as the output of the stage, since the deepest layer contains the strongest features.
%the deepest layer of each stage should contain the strongest features. 
We denote the output of conv$i$ as $C_i$ for i=2,3,4,5.
%The black dotted box in Fig. \ref{PFH_FPN_details} indicates the FPM.
%2, conv3, conv4, and conv5  
%in Table \ref{architecture}
%as $C_2$, $C_3$, $C_4$, and $C_5$, respectively.
%conv2, conv3, conv4, and conv5 outputs. 
%As there are totally five stages in ResNet, up to five feature maps can be used.

%There are often many layers producing output maps of the same size and we say these layers are in the same network stage. For our feature pyramid, we define one pyramid level for each stage. We choose the output of the last layer of each stage as our reference set of feature maps, which we will enrich to create our pyramid. This choice is natural since the deepest layer of each stage should have the strongest features.

%The black dotted box in Fig. \ref{PFH_FPN_details} indicates the FPM. %that constructs our top-down feature maps. 

The procedure of the proposed approach is as follows: (1) Using bilinear interpolation or transposed convolution, we upsample lower-resolution feature maps from higher stages by a factor of 2. That is, the top-down pathway hallucinates higher resolution features by upsampling low-resolution feature maps, but with more speaker-discriminative information, from higher stages.
(2) The upsampled feature maps are then enhanced with features from the bottom-up pathway via lateral connections.
Concretely, the top-down feature map is merged with the corresponding bottom-up feature map by element-wise addition.
Before merging, a $1\times1$ convolution in the lateral connection 
%is used to 
reduces the channel dimension of the bottom-up feature map to 32 which is the channel dimension of the lowest stage.
% U-Net 인용하면서 이것이 skip connection과 비슷한 효과로 정보를 보강해준다는 것을 설명.
These lateral connections play the same role as the skip connections in U-Net \cite{Ronneberger2015}. They directly transfer the high-resolution information from the bottom-up pathway to the top-down pathway.
%In this way, top-down feature maps are enhanced with features from the bottom-up pathway via lateral connections.
(3) This process is repeated from the top stage to the bottom stage. 
At the beginning, a $1\times1$ convolutional layer reduces the channel dimension of $C_5$ to 32. 
%At the beginning of the iteration, we add a $1\times1$ convolutional layer on $C_5$ to reduce the channel dimension to 32. 
%we add a $1\times1$ convolutional layer on $C_5$ to reduce the channel dimension to 32.
%produce the lowest resolution feature map. 
(4) Finally, an additional convolutional layer is appended to each merged feature map to reduce the aliasing effect of upsampling. This final set of feature maps is called $\{P_2, P_3, P_4, P_5\}$ corresponding to $\{C_2, C_3, C_4, C_5\}$ that are respectively of the same time-frequency resolution. 
%As the 
% 그림에서 c0 까지 그려야됨... P_1은 여기서 사용안했으므로 굳이..
%In the case of $P_1$,  

The FPM combines lower-resolution features with higher-level speaker information and higher-resolution features with lower-level speaker information.
%via a top-down pathway and lateral connections.
The result is a feature pyramid that has rich speaker information at all stages. 
In other words, the FPM plays the role of feature enhancement for MSA. Furthermore, the FPM reduces the total number of parameters in the network because the number of channels at stage 2, 3, and 4 are reduced by the $1\times1$ convolution in lateral connections.
%as the lateral connections reduce the number of channels at stage 2, 3, and 4, the total number of parameters in the network is decreased.
% 이 부분에서 Residual Networks Behave Like Ensembles of Relatively Shallow Networks 인용. MLA의 여러 path를 이용하는 것이 residual path의 ensemble 효과?  그런데 input acoustic으로부터 다양한 resolution의 feature들을 뽑고,  이것을 aggergation하여 ensemble로 이용하므로 짧은 발화 및 긴 발화 모두에 대해 잘 동작한다. 

A recent study shows that the collection of paths having different lengths in ResNet exhibits ensemble-like behavior \cite{Veit2016}. Similarly, we can interpret that the MSA method uses an ensemble of multi-scale features from different paths.
As the variable-length feature maps are used to extract speaker embeddings,
we expect that speaker verification performance will be improved for variable-duration test utterances, especially with the FPM. In our experiments, we show that the MSA with FPM provides improved performance for both short and long utterances.
%outperforms baseline systems for both short and long utterances.

\section{Experimental setup}
\label{sec:typestyle}

\subsection{Datasets}
We use the VoxCeleb1 \cite{Nagrani2017} and VoxCeleb2 \cite{Chung2018} datasets. Both are for large scale text-independent speaker recognition, containing 1,250 and 5,994 speakers, respectively. 
The utterances are extracted from YouTube videos where the speech segments are corrupted with real-world noise.
%The speech segments are corrupted with real world noise.
Both datasets are split into development (dev) and test sets.
%We adopt the same strategy as that in \cite{Chung2018} for evaluation. 
%In particular, 
The dev set of VoxCeleb2 is used for training and the test set of VoxCeleb1 is used for testing. 
There are no overlapping speakers between them.

When evaluating the performance on short utterances, our test recordings are cut into four different durations: 1 s, 2 s, 3 s, and 5 s, which is determined by the energy-based voice activity detection (VAD). If the length of the utterance is less than the given duration, the entire utterance is used.

\subsection{Implementation details}
The input features are 64-dimensional log Mel-filterbank features with a frame-length of 25 ms, which are mean-normalized over a sliding window of up to 3 s. 
Neither VAD nor data augmentation is used for training.
In training, the input size of the ResNet is $64 \times 300$ for 3 s of speech
(i.e., $d=64$ and $T=300$ in Fig. \ref{PFH_FPN_details}). 
In testing, the entire utterance is evaluated at once.
%in the testing stage. 
The 128-dimensional speaker embeddings are extracted from the network. 
We report the equal error rate (EER) in \% and the minimum detection cost function ($C_{det}^{min}$) 
%($C_{det}^{min}$)
with the same settings as in \cite{Chung2018}. 
%at $P_{target}$ = 0.01 with the same settings as in \cite{Chung2018}. 
Verification trials are scored using cosine distance. 

The models are implemented using PyTorch \cite{Paszke2017} and optimized by stochastic gradient descent with momentum 0.9. The mini-batch size is 64, and the weight decay parameter is 0.0001.
%is set to 0.0001. 
%All models are trained on a single NVIDIA GeForce GTX 1080 Ti GPU.
We use the same learning rate (LR) schedule as in \cite{Cai2018}, with the initial LR of 0.1.
For LDE, we use the same method as in \cite{Jung2019}. 
Before the LDE layer, a $1\times1$ convolution is applied to change the number of channels to 64. After the LDE layer, $L_2$-normalization and an FC layer are added to reduce the dimension to 128. The number of codewords is 64.
%set to 64.
When applying MSA, both the LDE and FC layers are shared by all stages.
On the other hand, the parameters of the SAP layers are not shared.

\begin{table}[t]
\centering
\renewcommand{\arraystretch}{1.0}
\caption{Comparison of Single, MSFA, and MSEA. The softmax loss and GAP are used for all systems.
In this paper, Single denotes using only single-scale features (Fig. \ref{PFH_FPN}(a)), FPM-B and FPM-TC are the FPMs with bilinear upsampling and transposed convolution upsampling, respectively.}
%The results of the proposed approaches are shown in bold.}
%The numbers in bold indicate the best results.}
%is the FPM with bilinear upsampling, and FPM-TC is the FPM with transposed convolution upsampling. 
%The results of the proposed approaches are shown in bold.}
%``\# params" : the number of parameters.}
\vspace{-0.25cm}
\label{MSA_comparison}
\begin{scriptsize}
\begin{tabular}{c|c||c|c|c}
\hline
\multicolumn{2}{c||}{Systems}               & $C_{det}^{min}$ & EER (\%)  & \# Parameters \\ \hline
\multicolumn{2}{c||}{Single}                & 0.423            & 4.55     & 5.77M         \\ \hline
\multirow{3}{*}{MSFA} & w/o FPM            & 0.437            & 4.30        & 6.20M         \\ %\cline{2-5} 
                      & \textbf{w/ FPM-B}   & \textbf{0.398}   & 4.22 & \textbf{5.82M}          \\ %\cline{2-5} 
                      & \textbf{w/ FPM-TC}  & 0.408   & \textbf{4.01} & 5.85M          \\ \hline
\multirow{3}{*}{MSEA} & w/o FPM            & 0.416            & 4.22          & 5.90M          \\ %\cline{2-5} 
                      & \textbf{w/ FPM-B}   & \textbf{0.403}   & 4.20 & \textbf{5.83M}          \\ %\cline{2-5} 
                      & \textbf{w/ FPM-TC}  & 0.411   & \textbf{4.01} & 5.85M          \\ \hline
\end{tabular}        
\vspace{-0.4cm}
\end{scriptsize}
\end{table}

\section{Results}
\label{sec:majhead}

\subsection{Performance in different MSA methods}

%We use the softmax loss (SM) and global average pooling (GAP) for all cases. 
In Table \ref{MSA_comparison} and \ref{poolings_comparison}, the models are trained on the VoxCeleb1 dev set.
Table \ref{MSA_comparison} compares with (w/) and without (w/o) the FPM in two MSA methods.
%two MSA approaches with and without the FPM.
%using only single-scale features (Single), MSFA, and MSEA. 
%two MSAs with and without the FPM.
%The system using only single scale features (Single) shows worst performance with the smallest number of parameters (5.77M).
MSFA w/o FPM and MSEA w/o FPM correspond to the approaches in Gao \textit{et al.} \cite{Gao2019} and Seo \textit{et al.} \cite{Seo2019}, respectively.
%, with the same architecture and loss function.
In both MSA approaches, we aggregate feature maps from 3 different stages: $C_3$, $C_4$, and $C_5$ for w/o FPM and $P_3$, $P_4$, and $P_5$ for w/ FPM.
%aggregate features from stage 2, 3, and 4 (i.e., $\{C_3, C_4, C_5\}$ for w/o FPM or $\{P_3, P_4, P_5\}$ for w/ FPM).
Note that the output of stage $i$ is $C_{i+1}$ for $i=1,2,3,4$ because the stage 1 corresponds to conv2 as shown in Table \ref{architecture}.
We can see that both MSAs yield better performance than Single, but with more parameters.

As we discussed in Section \ref{sec:priorworks}, the MSFA uses upsampling so that the three feature maps have the same spatial size.
In this work, bilinear upsampling is applied to $C_5$ since using transposed convolution does not improve the performance of MSFA with additional parameters.
%Here, bilinear interpolation is used for the system without the FPM (w/o FPM) as using transposed convolution does not improve the performance. 
%As using transposed convolution does not improve the performance, we use bilinear upsampling.
On the other hand, for the FPM, both bilinear and transposed convolution upsampling are used in the top-down pathway, which correspond to FPM-B and FPM-TC, respectively.
%for upsampling in the top-down pathway (FPM-B and FPM-TC, respectively). 
Among the three systems, w/o FPM has the worst performance with the largest number of parameters. 
%(EER=4.30\% and $C_{det}^{min}$=0.437) with the largest number of parameters. 
By adding the FPM, we obtain better performance with fewer parameters. 
The number of parameters is reduced because the channel dimension of feature maps at selected stages is reduced as we discussed in Section \ref{sec:FPM_section}.
%by $1\times1$ convolutions. 
FPM-TC has slightly more parameters than FPM-B because the transposed convolutional layer has learnable parameters.
FPM-B achieves a relative improvement of 8.92\% in $C_{det}^{min}$ over w/o FPM. FPM-TC shows a relative improvement of 6.74\% in EER over w/o FPM.

For MSEA, FPM-B provides the best $C_{det}^{min}$ (0.403), and FPM-TC obtains the best EER (4.01\%). 
%The FPM improves performance of both MSFA and MSEA.
%Both MSFA and MSEA improve performance by applying the proposed FPM. 
%The performances of both MSFA and MSEA are 
The proposed FPM improves the performance of both MSFA and MSEA, achieving similar performance with a similar number of parameters. Therefore, we only use MSEA in the following experiments. Moreover, the MSEA is more flexible to use various number of stages unlike the MSFA which is developed to use only 3 stages.
%As the MSFA is developed to use only 3 stages, it is not proper to extend for the MSFA to utilize more stages. The MSEA is more flexible to be extended to use various stages.

\begin{table}[t]
\centering
\renewcommand{\arraystretch}{1.00}
\caption{Performance comparison with and without the FPM for three pooling strategies: GAP, SAP \cite{Cai2018}, and LDE \cite{Cai2018}. MSEA with softmax loss is applied for all the systems.}
\vspace{-0.25cm}
\label{poolings_comparison}
\begin{scriptsize}
%\begin{tabular}{c|c|c|c}
%\begin{adjustbox}{max width=\textwidth}
\begin{tabular}{c||c|c|c|c|c|c}
\hline
\multirow{2}{*}{Systems}    & \multicolumn{2}{c|}{GAP} & \multicolumn{2}{c|}{SAP} & \multicolumn{2}{c}{LDE} \\ \cline{2-7} 
                         & $C_{det}^{min}$    & EER  & $C_{det}^{min}$    & EER        & $C_{det}^{min}$      & EER        \\ \hline
Single         & 0.423         & 4.55     & 0.410         & 4.38     & 0.421         & 4.44     \\ %\hline
w/o FPM                    & 0.416         & 4.22     & 0.416         & 4.24     & 0.435         & 4.09     \\ %\hline
\textbf{w/ FPM-B}                 & \textbf{0.403}         & 4.20     & \textbf{0.393}         & 4.13     & 0.402         & 3.84     \\ %\hline
\textbf{w/ FPM-TC}                & 0.411         & \textbf{4.01}     & 0.408         & \textbf{4.09}     & \textbf{0.368}         & \textbf{3.63}     \\ \hline
\end{tabular}
%\end{adjustbox}
\end{scriptsize}
\vspace{-0.15cm}
\end{table}

\begin{table}[t]
\centering
\renewcommand{\arraystretch}{1}
\caption{Comparison with state-of-the-art systems. In the parentheses, the first and second terms are the used loss function and pooling layer. For the loss function, C, ASM, SM, and EAMS denote contrastive, A-softmax \cite{Liu2017}, softmax, and EAM softmax \cite{Yu2019} loss, respectively. For the pooling layer, SPE, ASP, and SP denote spatial pyramid encoding \cite{Jung2019}, attentive statistics pooling \cite{Okabe2018}, and statistics pooling\cite{Snyder2018}, respectively.
* means that data augmentation is used and NR is ``not reported".}
\vspace{-0.25cm}
\label{comparison_SOTA}
\begin{scriptsize}
\begin{tabular}{c||c|c|c}
\hline
Systems              & Train set & $C_{det}^{min}$  & EER  \\ \hline
\textit{i}-vectors+PLDA \cite{Nagrani2017}    & Vox1  & 0.73 & 8.8     \\ 
VGG-M (C+GAP) \cite{Nagrani2017}     & Vox1  & 0.71 & 7.8     \\ 
ResNet-34 (ASM+SAP) \cite{Cai2018}   & Vox1  & 0.622 & 4.40    \\ 
ResNet-34 (ASM+SPE) \cite{Jung2019}  & Vox1  & 0.402 & 4.03     \\ 
TDNN (SM+ASP) \cite{Okabe2018}       & Vox1*  & 0.406 & 3.85     \\ 
\textbf{ResNet-34 (ASM+GAP) w/ FPM-TC}  & Vox1  & 0.393 & 3.52    \\ 
\textbf{ResNet-34 (ASM+LDE) w/ FPM-TC}  & Vox1  & \textbf{0.350} & \textbf{3.22}    \\ \hline
%ResNet-50 (C+GAP) \cite{Chung2018}   & Vox2  & 0.429 & 3.95     \\ 
Thin ResNet-34 (SM+GhostVLAD) \cite{Xie2019} & Vox2 & NR & 3.22  \\
ResNet-50 (EAMS+GAP) \cite{Yu2019}   & Vox2  & 0.278 & 2.94     \\ 
ResNet-34 (ASM+SPE) \cite{Jung2019}  & Vox2  & 0.245 & 2.61     \\ 
DDB+Gate (SM+SP) \cite{Jiang2019}  & Vox1\&2  & 0.268 & 2.31     \\ 
\textbf{ResNet-34 (ASM+GAP) w/ FPM-TC} & Vox2  & 0.228 & 2.17     \\ 
\textbf{ResNet-34 (ASM+LDE) w/ FPM-TC}  & Vox2  & \textbf{0.205} & \textbf{1.98}     \\ \hline
\end{tabular}
\end{scriptsize}
\vspace{-0.45cm}
\end{table}

\begin{table}[t]
\centering
\renewcommand{\arraystretch}{1}
\caption{EER (\%) of systems on the 5 s enrollment set.}
\vspace{-0.28cm}
\label{short_five}
\begin{scriptsize}
%\begin{tabular}{c|c|c|c}
\begin{tabular}{c||ccccc}
\hline
Systems                             & 1 s & 2 s & 3 s & 5 s & full \\ \hline
TDNN (ASM+SP)                       &  10.93  & 6.12 & 4.50  &  3.83  & 3.35      \\
TDNN (ASM+ASP)                      & 10.21   &  5.62  & 4.47 &  3.62  & 3.29     \\
ResNet-34 (ASM+SPE) \cite{Jung2019} &  12.13  & 5.60   & 4.10   & 3.45   & 3.07      \\
MSEA w/o FPM (ASM+GAP)              &  6.65  & 4.05   & 3.23   & 2.74   & 2.56      \\
\textbf{Proposed 1}                 &  \textbf{6.25}  &  3.95  & 2.99  &  2.54 &  2.35      \\
\textbf{Proposed 2}                 &  6.54 &  \textbf{3.76}  &  \textbf{2.80}  & \textbf{2.47} & \textbf{2.23}    \\ \hline
\end{tabular}
\end{scriptsize}
\vspace{-0.15cm}
\end{table}

\begin{table}[t]
\centering
\renewcommand{\arraystretch}{1}
\caption{EER (\%) of systems on the full-length enrollment set.}
\vspace{-0.28cm}
\label{short_full}
\begin{scriptsize}
%\begin{tabular}{c|c|c|c}
\begin{tabular}{c||ccccc}
\hline
Systems                             & 1 s & 2 s & 3 s & 5 s & full \\ \hline
TDNN (ASM+SP)                      & 9.92  & 5.50 & 4.06   & 3.47   & 3.04      \\
TDNN (ASM+ASP)                      &  9.63  &  5.02  & 3.87 &  3.36  & 2.92     \\
ResNet-34 (ASM+SPE) \cite{Jung2019} &  11.12  & 4.93    & 3.55   & 2.98   & 2.61      \\
MSEA w/o FPM (ASM+GAP)              &  6.13  & 3.68   &  2.84   & 2.50   & 2.31      \\
\textbf{Proposed 1}                 &  \textbf{5.85}  &  3.64  &  2.83  & 2.41 &  2.17      \\
\textbf{Proposed 2}                &  5.92  & \textbf{3.38} &  \textbf{2.54}  &  \textbf{2.17}  &  \textbf{1.98}    \\ \hline
\end{tabular}
\end{scriptsize}
\vspace{-0.4cm}
\end{table}

\subsection{Performance in different pooling methods}
In Table \ref{poolings_comparison}, we compare the performance among the three pooling strategies.
%Here, we use the softmax loss and apply the MSEA for all cases.
For the SAP layer, FPM-B provides the best $C_{det}^{min}$ (0.393), and FPM-TC obtains the best EER (4.09\%). 
%the results are similar to those for the GAP layer. 
%That is, FPM-B provides the best $C_{det}^{min}$ (0.393), and FPM-TC obtains the best EER (4.09\%). 
For the LDE layer, we only aggregate features from stage 3 and 4 because using features from stage 2 does not improve the performance.
FPM-TC achieves the best performance on both $C_{det}^{min}$ (0.368) and EER (3.63\%). For all the three pooling strategies, the FPM improves the performance of the MSA method.
We can see that the FPM is most effective for the LDE layer.

\subsection{Comparison with recent methods}
In Table \ref{comparison_SOTA}, we compare the two proposed systems with recently reported SV systems. 
Both proposed systems apply the MSEA with FPM-TC, and the combination of A-softmax and ring loss with the same settings as in \cite{Jung2019}. 
When we use the VoxCeleb2 dataset for training, we extract 256-dimensional speaker embeddings.
In the first system (Proposed 1), we aggregate features from all the stages using GAP.
In the second system (Proposed 2), we aggregate features from stage 2, 3, and 4 using LDE.
The proposed systems outperform other baseline systems in terms of both $C_{det}^{min}$ and EER.
%on both datasets.
Proposed 2 achieves the best performance among all the systems.
Using the VoxCeleb1 dataset for training, Proposed 2 obtains a $C_{det}^{min}$ of 0.350 and an EER of 3.22\%. Using the VoxCeleb2 dataset for training, Proposed 2 achieves a $C_{det}^{min}$ of 0.205 and an EER of 1.98\%.

%\subsection{Performance for different test durations}
In Table \ref{short_five} and \ref{short_full}, we evaluate the performance of several systems in 5 s and full-length enrollment conditions, respectively. For each condition, we evaluate the performance on five different test durations: 1 s, 2 s, 3 s, 5 s, and original full-length (full). The average duration of `full' is 6.3 s. All the results are obtained by our own implementation and VoxCeleb2 dataset is used for training. For TDNN, we follow the same architecture as in \cite{Okabe2018}. For all the baseline systems, we use the same acoustic features and optimization as our proposed systems.

Among baselines, the ResNet-based system using SPE (the advanced version of LDE) outperforms TDNN-based systems except for the 1 s test condition. Similarly, Proposed 2 achieves better results than Proposed 1 for utterances longer than 1 s. 
From these, we can see that LDE-based pooling shows a greater performance degradation on very short utterances than the other pooling strategies. 
MSEA w/o FPM is the approach in the 5th row of Table \ref{MSA_comparison}.
%, and Proposed 1 is MSEA with FPM-TC. 
We find that applying the FPM improves the performance of MSA for variable-duration test utterances. 
When we compare proposed systems with other state-of-the-art baseline systems including TDNN (x-vector) and ResNet-based systems, we also observe that the MSA with FPM achieves higher performance for both short and long utterances. 

\section{Conclusions}

In this study, we proposed a novel MSA method for TI-SV using a FPM.
We applied the FPM to two types of MSA methods, MSFA and MSEA. It enhances speaker-discriminative information on multi-scale features at multiple layers of a speaker feature extractor. 
%The FPM enhances speaker information on features at all levels of the frame-level feature extractor.
On the VoxCeleb dataset, experimental results showed that the FPM improves both MSA methods with fewer parameters, and works well with three popular pooling layers. The proposed systems obtained better results for both short and long utterances than the state-of-the-art baseline systems.

\section{Acknowledgements}

This material is based upon work supported by the Ministry of Trade, Industry and Energy (MOTIE, Korea) under Industrial Technology Innovation Program (No.10063424, Development of distant speech recognition and multi-task dialog processing technologies for in-door conversational robots).

\bibliographystyle{IEEEtran}

\bibliography{mybib}

% \begin{thebibliography}{9}
% \bibitem[1]{Davis80-COP}
%   S.\ B.\ Davis and P.\ Mermelstein,
%   ``Comparison of parametric representation for monosyllabic word recognition in continuously spoken sentences,''
%   \textit{IEEE Transactions on Acoustics, Speech and Signal Processing}, vol.~28, no.~4, pp.~357--366, 1980.
% \bibitem[2]{Rabiner89-ATO}
%   L.\ R.\ Rabiner,
%   ``A tutorial on hidden Markov models and selected applications in speech recognition,''
%   \textit{Proceedings of the IEEE}, vol.~77, no.~2, pp.~257-286, 1989.
% \bibitem[3]{Hastie09-TEO}
%   T.\ Hastie, R.\ Tibshirani, and J.\ Friedman,
%   \textit{The Elements of Statistical Learning -- Data Mining, Inference, and Prediction}.
%   New York: Springer, 2009.
% \bibitem[4]{YourName17-XXX}
%   F.\ Lastname1, F.\ Lastname2, and F.\ Lastname3,
%   ``Title of your INTERSPEECH 2020 publication,''
%   in \textit{Interspeech 2020 -- 20\textsuperscript{th} Annual Conference of the International Speech Communication Association, September 15-19, Graz, Austria, Proceedings, Proceedings}, 2020, pp.~100--104.
% \end{thebibliography}

\end{document}